\pdfoutput=1
\documentclass[journal,10pt,draftclsnofoot,onecolumn,a4paper]{IEEEtran}

\usepackage{caption}
\usepackage[pdftex]{graphicx}
\DeclareGraphicsExtensions{.pdf,.jpeg,.png,.jpg}
\usepackage{cite}
\usepackage{amsmath}
\usepackage{amssymb}					%Additional math symbols
\usepackage{stfloats}
\usepackage{url}
\usepackage{flushend}
\usepackage[latin1]{inputenc}
\usepackage{multirow}
\usepackage{pifont}
\usepackage{color}

\usepackage{enumerate}
\usepackage{siunitx}
\usepackage{breakurl}
\usepackage{epstopdf}
\usepackage{pbox}
\usepackage{booktabs} % Publication-quality tables
\usepackage{array, tabularx}
\usepackage{threeparttable} % Including footnotes in tables

\usepackage[hidelinks]{hyperref}

\usepackage{bm} % Bold greek symbols for math environments
\bmdefine{\bsigma}{\sigma}
\newcommand{\etal}{}% To make sure that \etal isn't already defined    
\def\etal/{\textit{et al.}}

\newcolumntype{Y}{>{\centering\arraybackslash}X}
\newcommand{\bmat}[1]{\begin{bmatrix}#1\end{bmatrix}}
\begin{document}
\sloppy
\pagenumbering{gobble}

\title{Model Reduction of Converter-Dominated Power Systems by Singular Perturbation Theory}

%\title{Singular Perturbation Method for the Model Order Reduction of Power Systems with a High Penetration of Power Converters}

\author{
	\vskip 1em
	Umberto Biccari\textsuperscript{a,b}, 
	Noboru Sakamoto\textsuperscript{a,c}, 
	Eneko Unamuno\textsuperscript{d}*, 
	Danel Madariaga\textsuperscript{e}, 
	Enrique Zuazua\textsuperscript{a,f,g}, 
	Jon Andoni Barrena\textsuperscript{d} 
	
	\thanks{The work of all authors was partially supported by the ELKARTEK project KK-2018/00083 ROAD2DC of the Basque Government. In addition, the work of U. B., E. U., E. Z., and J. A. B. was partially supported by the Grant MTM2017-92996-C2-1-R/2-R COSNET of MINECO (Spain). This project has also received funding from the European Research Council (ERC) under the European Union's Horizon 2020 research and innovation programme (grant agreement NO. 694126-DyCon). The work of E. Z. has been partially supported by the Alexander von Humboldt-Professorship program, the European Union's Horizon 2020 research and innovation programme under the Marie Sklodowska-Curie grant agreement NO. 765579-ConFlex and by the grant ICON-ANR-16-ACHN-0014 of the French ANR. The work of U. B. and E. Z. was partially supported by the Air Force Office of Scientific Research (AFOSR) under Award NO. FA9550-18-1-0242. 
		
		*Corresponding author (\href{mailto:eunamuno@mondragon.edu}{e-mail: eunamuno@mondragon.edu})
		
		\textsuperscript{a} Chair of Computational Mathematics, Fundaci\'on Deusto Av. de las Universidades 24, 48007 Bilbao, Basque Country, Spain.
		
		\textsuperscript{b} Facultad de Ingenier\'ia, Universidad de Deusto, Avenida de las Universidades 24, 48007 Bilbao, Basque Country, Spain.
		
		\textsuperscript{c} Faculty of Science and Engineering, Nanzan University, Nagoya, 466-8673, Japan
		
		\textsuperscript{d} Electronics and Computing Department, Mondragon Unibertsitatea, Arrasate-Mondrag\'on, 20500 Spain.
		
		\textsuperscript{e} Ingeteam R\&D Europe S. L., 48170 Zamudio, Spain.
		
		\textsuperscript{f} Chair in Applied Analysis, Alexander von Humboldt-Professorship, Department of Mathematics Friedrich-Alexander-Universit\"at Erlangen-N\"urnberg, 91058 Erlangen, Germany.
		
		\textsuperscript{g} Departamento de Matem\'aticas, Universidad Aut\'onoma de Madrid, 28049 Madrid, Spain.
		
	}
}

\maketitle
\begin{abstract}
	The increasing integration of power electronic devices is driving the development of more advanced tools and methods for the modeling, analysis, and control of modern power systems to cope with the different time-scale oscillations. In this paper, we propose a general methodology based on the singular perturbation theory to reduce the order of systems modeled by ordinary differential equations and the computational burden in their simulation. In particular, we apply the proposed methodology to a simplified power system scenario comprised of three inverters in parallel---controlled as synchronverters---connected to an ideal grid. We demonstrate by time-domain simulations that the reduced and decoupled system obtained with the proposed approach accurately represents the dynamics of the original system because it preserves the non-linear dynamics. This shows the efficiency of our technique even for transient perturbations and has relevant applications including the simplification of the Lyapunov stability assessment or the design of non-linear controllers for large-scale power systems.
\end{abstract}


%\markboth{IEEE JOURNAL OF EMERGING AND SELECTED TOPICS IN POWER ELECTRONICS}%
%{}

\definecolor{limegreen}{rgb}{0.2, 0.8, 0.2}
\definecolor{forestgreen}{rgb}{0.13, 0.55, 0.13}
\definecolor{greenhtml}{rgb}{0.0, 0.5, 0.0}
	
% \newpage
\section{Introduction}

The massive inclusion of power electronic devices in power systems driven by the integration of distributed generation, energy storage components and the substitution of classical interconnection equipment---such as transformers or tie lines---is significantly changing the way in which power grids are studied and operated. 

The ever-increasing complexity of modern electrical networks results in the more and more frequent appearance of novel phenomena, such as diverse timescale oscillations caused by the faster dynamic behavior of power electronic converters connected to the system.

For these reasons, many tools and techniques currently employed to model, analyze, and control electrical systems are becoming obsolete. Consequently, more advanced control and management approaches, as well as novel analysis and modeling instruments, are required to cope with challenges such as the high-scale of power systems, the lack of primary reserve or inertial response, the interactions between devices that might bring the system to instability, power quality issues due to resonances and unbalanced grid conditions, etc. \cite{Milano2018}.

In the wide spectrum of different methodologies developed for the treatment of power systems, Model Order Reduction (MOR) techniques have played an important role. One of the main reasons is that they generally facilitate the study of the dynamics and interactions of the elements connected to the grid and, more in general, the overall analysis of complex power systems by keeping only the most relevant information but, at the same time, accurately representing the complete phenomena. The aim of these techniques is thus to reduce the computational burden for the simulation of high-order models, but also to simplify the study of different properties such the system's stability, e.g. by determining Lyapunov functions for the estimation of the regions of attraction (see  \cite[Chapter 7]{kokotovic1999singular} or \cite{son2017solving}).

Some of the most relevant MOR techniques for linear systems have been discussed and compared in \cite{Ramlal2016}. However, in most cases the non-linear nature of power electronic devices requires the adaptation of these techniques to preserve the non-linear behavior also in the reduced models. 

In the context of power systems, some classical MOR approaches are based on multiple-scale analysis \cite{kevorkian2012multiple}, according to which the fast (short timescale) and slow (long timescale) dynamics of a system are almost independent provided that their timescales are sufficiently detached. According to this principle, in a multiple-scale system, it is possible to separate the state variables into those contributing to the fast and to the slow dynamics. The fast states can be replaced with a pure gain when evaluating slow dynamics, whereas the slow states can be held stationary when evaluating fast dynamics. This yields reduced models with lower order and narrower timescale than the full model \cite{kokotovic1976singular}. 

In this framework, one of the most classical MOR approaches is the so-called modal truncation, in which the less relevant modes of the system are truncated, and the dynamics is represented only with the dominant modes. This is the approach adopted for instance in \cite{Gu2018}, where Gu \etal/ propose a MOR method based on the modal truncation that keeps the non-linear response of the original system to improve the accuracy of the reduced models. 

Alternatively, Kodra \etal/ carry out in \cite{Kodra2016} a MOR inspired on the Singular Perturbation (SP) theory, based on the seminal works of Kokotovic \etal/ (see \cite{kokotovic1999singular}), for the analysis of an islanded micro-grid and they discuss advantages and disadvantages of this procedure when applying it to different scenarios. More specifically, they point out that the standard SP theory which separates slow and fast subsystems based on the values of eigenvalue real parts leads poor approximations due to highly oscillatory nature of the system. 

This is one of the motivations of the present paper and 
we analyze an ODE model describing the dynamics of a power system with a high penetration of power electronic converters, composed by three inverters with $\mathit{LC}$ filters connected in parallel to a common grid. Our final aim is to define a methodology based on the linearization of the system around a steady-state to identify the slow and highly oscillatory states and the application of singular perturbation theory to carry out the MOR of the mentioned model by carrying out a block diagonalization of the equations. Our methodology, corroborated by dynamic time-domain simulations, provides an easily applicable approach to suitably identify and decouple fast oscillation and slow dynamic modes. Moreover, even if the initial steps of the methodology rely on the identification of the eigenvalues from a small-signal model, the final representation in our approach has the advantage of maintaining the non-linear nature of the original model, being therefore particularly suitable for a global analysis.

The rest of the paper is structured as follows: Section~\ref{sec:two} introduces the proposed model order reduction methodology generalized for any dynamical system. Section~\ref{sec:three} presents the main properties of the power system scenario in which the methodology is applied and describes the equations that represent its dynamic behavior. In Section~\ref{sec:four}, the validity of the proposed methodology is demonstrated by comparing the results of the reduced models to the original system of equations via time-domain simulation. In Section~\ref{sec:conc} we collect the most relevant conclusions of the study.

\section{Proposed methodology} \label{sec:two}

The methodology proposed in this paper has been developed to reduce the complexity of power systems with a high penetration of power converters and to facilitate their study. Notwithstanding, for the sake of completeness, we shall remark that the methodology we propose is not limited to the scenario considered in the present paper. In fact, it may be extended and applied to other types of dynamical systems (electrical, thermal, mechanical, etc.) modeled by ODEs.

Order reduction techniques are generally used to simplify high-order models into low-order approximations without having significant effects on the system's properties. 

In the context of micro-grids or specific embedded systems---such as generators, storage devices, and renewable energy sources for dynamic simulations or controller design purposes---several order reduction methods have been employed in the past. 

For example, in \cite{freitas2008gramian}, Gramian-based reduction methods have been applied for the MOR of large sparse power system descriptor models.

In \cite{chaniotis2005model}, the authors employ Krylov subspace techniques to simplify the complicated model of an \textit{m}-machine \textit{n}-bus power system that can be used in a microgrid environment. 

Finally, in \cite{luo2014spatiotemporal}, a model-reduction scheme is proposed based on SP and Kron reduction applied to large-signal dynamic models of inverter-based islanded microgrids. 

In this work we focus on the singular perturbation approach for the MOR of the power system we analyze. For completeness, we have to stress that from a purely mathematical perspective the singular perturbation technique that we employ is not new, being based on the classical theory developed, among others, by Kokotovic, Levin, and Levinson (see \cite{kokotovic1976singular,kokotovic1999singular,levin1954singular}). Nevertheless, to the best of our knowledge, it has not been applied yet in the context of our model.

\begin{figure}[h!]
	\centering
	\includegraphics[width=0.6\textwidth]{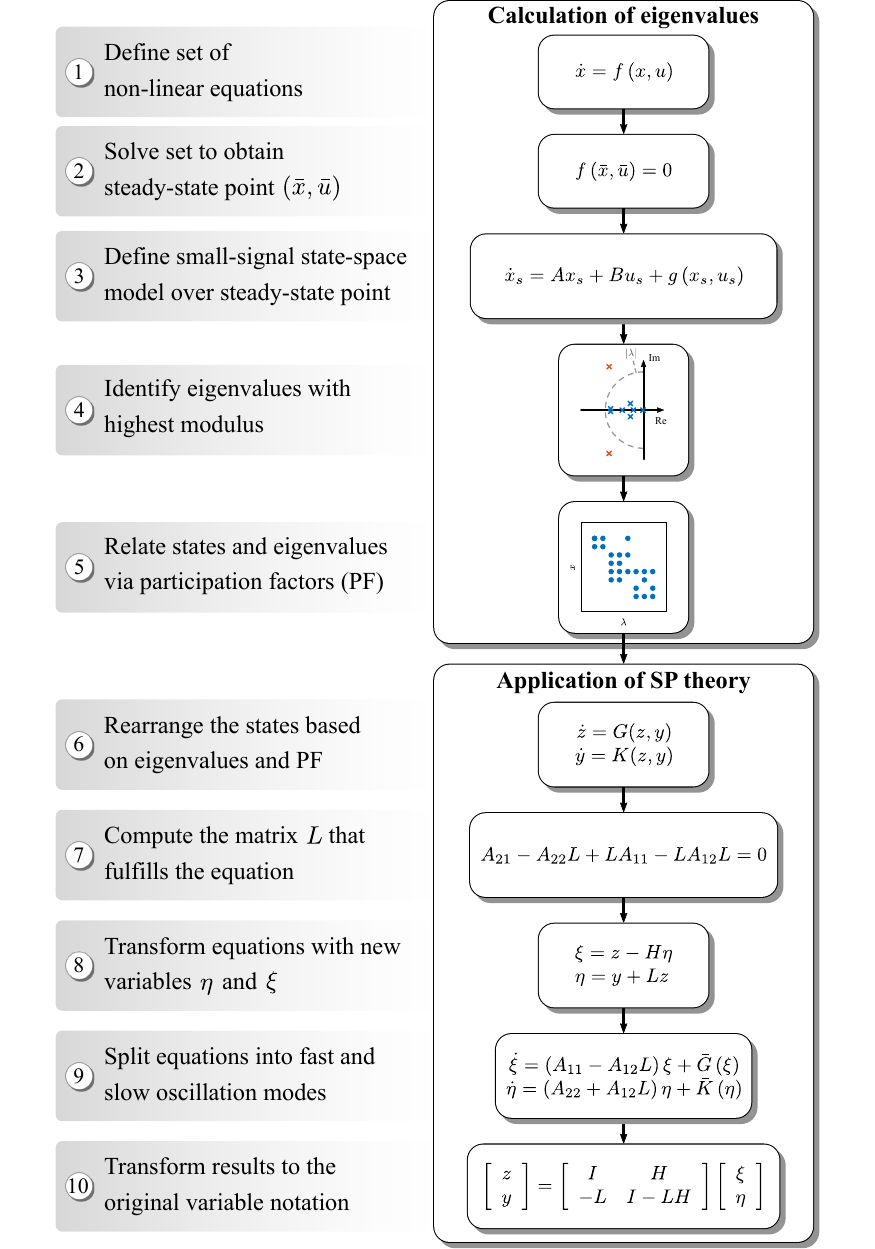}
	\caption{Simplified flowchart of the proposed MOR algorithm}
	\label{fig:methodology}
\end{figure}

Let us now give a general description of the proposed methodology. In order to give a more clear picture, the most relevant steps of our approach are represented in Figure~\ref{fig:methodology}, which can be separated in two main blocks.

\subsection{System's eigenvalues calculation}

In the first part of our methodology, the aim is to compute the eigenvalues at a specific point of operation from the small-signal model and to determine their relation with the model's states. To this end, we proceed in five steps:
\begin{enumerate}
	\item Define the non-linear ODEs that describe the dynamic behavior of our system;
	\item Compute the steady state for specific input values to obtain the operation point $\left(\bar x,\bar u\right)$;
	\item Obtain the small-signal model by linearizing the system around the operation point $\left(\bar x,\bar u\right)$; 
	\item Compute the eigenvalues of the small-signal model to identify the ones with the largest modulus; 
	\item Calculate the participation factors (PF) of these eigenvalues to identify how they are related to the model's states.
\end{enumerate}

\subsection{Application of singular perturbation theory}
The purpose in this second part is to reduce the order of the original system by decoupling the states with fast and slow dynamic.

The mathematical background at the basis of the following steps is described in Appendix~\ref{app:sp}.

Based on the eigenvalues analysis of the first part, the application of SP theory is carried out as follows: 
\begin{enumerate}
	\item[6)] Rewrite the equations in our model in the form \eqref{eq:SP_general1}-\eqref{eq:SP_general2}: the state $y$ and $z$ are associated to the eigenvalues with large and small modulus, respectively; 
	\item[7)] Compute the matrix $L$ solution of the equation \eqref{eq:L}; 
	\item[8)] Transform the dynamic equations through a two-steps change of variables from $\left[ {z,y} \right]^ \top$ to $\left[ {\xi, \eta} \right]^ \top$, employing the matrix $L$ previously computed and a new matrix $H$ solution of a Sylvester equation;
	\item[9)] Split the ODEs of the new system---which are represented with the new variables $\left[ {\xi, \eta} \right]^ \top$---into fast and slow states. As will be shown in the following sections, this representation can be easily employed to simulate the system as two completely isolated subsystems;
	\item[10)] Transform back the obtained results to the original variable notation by applying the inverse change of variables.
\end{enumerate}

As a final comment let us stress the fact that, although in steps 3--5 we passed through the small-signal system and its eigenvalue analysis, this is a preliminary step to rearrange the states of the system but the final representation obtained with our proposed methodology maintains the original non-linear nature of the model. 

This, on the one hand, has the advantage of keeping the original level of accuracy in the final decoupled system, since we do not lose information about the high order non-linear states. 

On the other hand, the fact that the non-linear features are preserved after the model reduction makes our approach particularly convenient for being applied to global stability analysis and, possibly, non-linear control design.

\section{Modeling of a simplified power system scenario} \label{sec:three}

\begin{figure}[h!]
	\centering
	\includegraphics[width=0.5\textwidth]{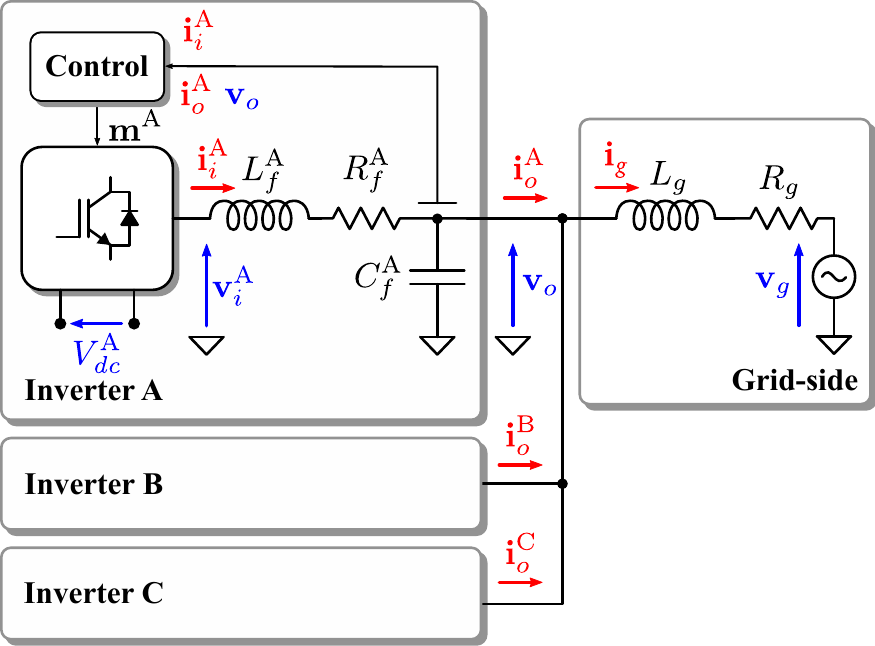}
	\caption{Studied power system scenario}
	\label{fig:analysedscenario}
\end{figure}

Figure~\ref{fig:analysedscenario} illustrates the scenario analyzed in this paper, which is comprised by three inverters with $\mathit{LC}$ filters connected in parallel to a common grid. In this case, the inverters are controlled to emulate the behavior of classical synchronous generators, contributing to the primary regulation and inertial response of the grid under frequency variations---also known as \textit{grid-forming} inverters. The control techniques that have been implemented in this case are called synchronverters, which have been widely studied in previous publications \cite{Zhong2011,Zhong2014,Zhong2017}.

The scenario defined to validate the proposed MOR methodology is a simplified version of a higher-scale power system with a high penetration of power converters. It is suitable because it represents the dynamics not only of the electrical part but also of the controllers of the converters and their interactions. 

The overall system can represented in the state-space domain with a set of ODEs in the form of $f\left( {x,u} \right)$, where $x$ denotes the state vector and $u$ represents the input vector. In this case $x$ and $u$ are time-dependent variables but for the sake of readability this dependency is not shown explicitly throughout the text. We must also point out that in the following sections all the variables are represented in the \textit{dq} domain in vector notation (e.g. ${{\bf{v}}_g} = {v_{{g_d}}} + j{v_{{g_q}}}$, $j$ being the imaginary operator) and are referred to the grid reference frame---i.e. aligned with the grid voltage ${{\bf{v}}_g}$---unless otherwise stated. In order to normalize the system, all the variables and parameters are represented in a per unit (p.u.) notation with $S_b$, $V_b$ and $\omega_b$ as the base power, base voltage and base frequency, respectively. The meaning of these variables is represented in Figure~\ref{fig:analysedscenario}.

\subsection{Grid-side}
The grid-side is modeled as an ideal voltage source in series with an $\mathit{RL}$ impedance, and its dynamic behavior can be represented by a first-order ODE describing the time-domain evolution of the grid-side current:
\begin{equation}
\frac{{d{{\bf{i}}_g}}}{{dt}} = {\omega _b}\left( {\frac{1}{{{L_g}}}\left( {{{\bf{v}}_o} - {R_g}{{\bf{i}}_g} - {{\bf{v}}_g}} \right) - j{\omega _g}{{\bf{i}}_g}} \right)
\end{equation}
where the grid voltage is an input to the system defined as ${{\bf{v}}_g} = V_g + j0$.

\subsection{Inverter filters}
The filters are modeled as $\mathit{RLC}$ impedances, and their dynamic behavior can be represented by the inductor currents and the capacitor voltages. 

The inverter currents can be represented as:
\begin{equation}
\frac{{d{\bf{i}}_i^k}}{{dt}} = {\omega _b}\left( {\frac{1}{{L_f^k}}\left( {{{\bf{m}}^k}V_{dc}^k - R_f^k{\bf{i}}_i^k - {{\bf{v}}_o}} \right) - j{\omega _g}{\bf{i}}_i^k} \right)
\end{equation}
where the superscript $k$ indicates the inverter's name ($\mathrm{A}$, $\mathrm{B}$ or $\mathrm{C}$) and $\mathbf{m}$ is the modulation index coming from the controller, which is described in Section~\ref{ssec:control}.

Regarding the voltages of the filter capacitors, they can be represented in a single equation by summing the inverter output currents and filter capacitors as follows:
\begin{equation}
\frac{{d{{\bf{v}}_o}}}{{dt}} = {\omega _b}\left( {\frac{1}{{{C_T}}}\left( {{{\bf{i}}_T} - {{\bf{i}}_g}} \right) - j{\omega _g}{{\bf{v}}_o}} \right)
\end{equation}
where ${C_T} = \sum {C_f^k}$ and ${i_T} = \sum {i_f^k}$.

\subsection{Control part}\label{ssec:control}
As mentioned previously, the controllers in this case are synchronverters that emulate the behavior of classical synchronous machines. A block diagram of this type of controller is represented in Figure 4. The variables in this figure and in the following equations are represented generally (without the name of the inverter), but the same control strategy is employed for the three inverters connected in parallel.

\begin{figure}[h!]
	\centering
	\includegraphics[width=0.5\textwidth]{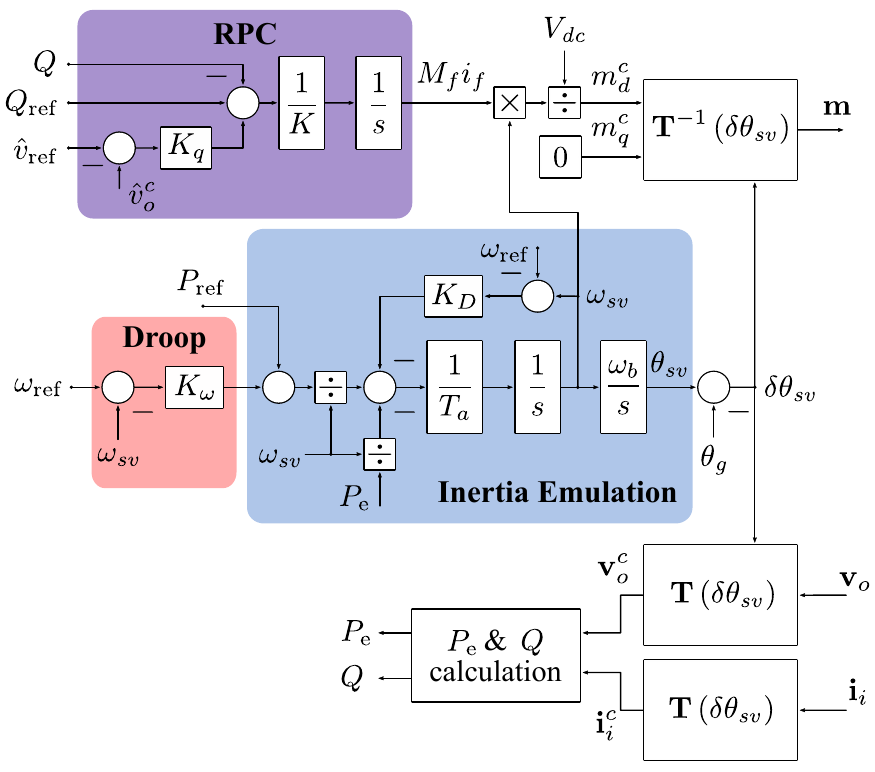}
	\caption{Block diagram of the synchronverter controller}
	\label{fig:synchronverter}
\end{figure}

With the superscript $c$ in variables such as the output voltage (${\bf{v}}_o^{{\mkern 1mu} c}$) and the inverter current  (${\bf{i}}_o^{{\mkern 1mu} c}$), we denote that the variable is referred to  the controller reference frame. The rotation between the grid and the controller reference frame is carried out with the following transformation:
\begin{equation}
\begin{array}{c}
{\bf{v}}_o^{{\mkern 1mu} c} = {\bf{T}}\left( {\delta {\theta _{sv}}} \right){{\bf{v}}_o}\\
{\bf{i}}_i^{{\mkern 1mu} c} = {\bf{T}}\left( {\delta {\theta _{sv}}} \right){{\bf{i}}_i}
\end{array}
\end{equation}
where ${\bf{T}}(\delta {\theta _{sv}})$ is a rotation matrix based on \cite{Rygg2017,Zhang2019}:
\begin{equation}
{\bf{T}}\left( {\delta \theta } \right) = \left[ {\begin{array}{*{20}{c}}
	{\cos \left( {\delta {\theta _{sv}}} \right)}&{\sin \left( {\delta {\theta _{sv}}} \right)}\\
	{ - \sin \left( {\delta {\theta _{sv}}} \right)}&{\cos \left( {\delta {\theta _{sv}}} \right)}
	\end{array}} \right]
\end{equation}

The reactive power controller can be represented by the equation:
\begin{equation}\label{eq:mfif}
\frac{{d\left( {{M_f}{i_f}} \right)}}{{dt}} = \frac{1}{K}\left[ {{Q_{{\rm{ref}}}} - Q + {K_q}\left( {{{\hat v}_{{\rm{ref}}}} - \hat v_o^c} \right)} \right].
\end{equation}	

The frequency ${\omega _{sv}}$ of the controller and the angle ${\theta _{sv}}$ are obtained from the inertia emulation part and can be written in the differential form as follows:
\begin{equation}\label{eq:wsv}
\frac{{d{\omega _{sv}}}}{{dt}} = \frac{1}{T_a}\left[ {\frac{{{P_m}}}{{{\omega _{sv}}}} - \frac{{{P_e}}}{{{\omega _{sv}}}} - {K_D}\left( {{\omega _{sv}} - {\omega _{{\rm{ref}}}}} \right)} \right]
\end{equation}
\begin{equation}
\frac{{d\delta {\theta _{sv}}}}{{dt}} = {\omega _b}\left( {{\omega _{sv}} - {\omega _g}} \right)
\end{equation}
where $\omega _g$ is the constant frequency of the grid. Similarly, in (\ref{eq:wsv}) $P_m$ is the emulated mechanical power of the converter, which is calculated by adding a power reference to the output of a frequency droop regulator. This can be expressed as:
\begin{equation}
{P_m} = {P_{{\rm{ref}}}} + {K_\omega }\left( {{\omega _{{\rm{ref}}}} - {\omega _{sv}}} \right)
\end{equation}
where ${P_{{\rm{ref}}}}$ and ${\omega _{{\rm{ref}}}}$ are active power and frequency references defined as inputs, respectively.

$P_e$ and $Q$ appearing in equation \eqref{eq:mfif} are calculated by considering that, in a balanced three-phase system, the average active and reactive powers can be directly obtained by multiplying the output current with the complex conjugate of the current as follows:
\begin{equation}
{P_e} + jQ = {\bf{v}}_o^c{\bf{i}}_i^{c*} = \left( {v_{{o_d}}^c + jv_{{o_q}}^c} \right)\left( {i_{{i_d}}^c - ji_{{i_q}}^c} \right)
\end{equation}

Moreover, ${\hat v_{{\rm{ref}}}}$ in (\ref{eq:mfif}) is a reference voltage, while $\hat v_o^{{\mkern 1mu} c}$ is the amplitude of the measured output voltage, which can be calculated as:
\begin{equation}
\hat v_o^{{\mkern 1mu} c} = \sqrt {{{\left( {v_{{o_d}}^{{\mkern 1mu} c}} \right)}^2} + {{\left( {v_{{o_q}}^{{\mkern 1mu} c}} \right)}^2}} 
\end{equation}

\subsection{Overall state and input vectors}
Based on the ODEs defined above, the state vector $x \in \mathbb{R}^{19}$ of the overall system is comprised by the following state variables:
\begin{equation}
{x} = {\left[ {\begin{array}{*{20}{c}}
		{{{\bf{v}}_o}}&{{{\bf{i}}_i^{\mathrm{A}}}}&{{{\bf{i}}_i^{\mathrm{B}}}}&{{{\bf{i}}_i^{\mathrm{C}}}}&{{{\bf{i}}_g}}&{{\omega _{sv}^{\mathrm{A}}}}&{\delta {\theta _{sv}^{\mathrm{A}}}}&{{M_f}{i_f}}^{\mathrm{A}}&{{\omega _{sv}^{\mathrm{B}}}}&{\delta {\theta _{sv}^{\mathrm{B}}}}&{{M_f}{i_f}}^{\mathrm{B}}&{{\omega _{sv}^{\mathrm{C}}}}&{\delta {\theta _{sv}^{\mathrm{C}}}}&{{M_f}{i_f}}^{\mathrm{C}}
		\end{array}} \right]^ \top }
\end{equation}
where the superscripts indicate the number of the converter.

Similarly, the input vector $u \in \mathbb{R}^{13}$ is defined with the external references:
\begin{equation}
{u} = {\left[ {\begin{array}{*{20}{c}}
		{V_g}&{{P_{{\rm{ref}}}^{\mathrm{A}}}}&{{Q_{{\rm{ref}}}^{\mathrm{A}}}}&{{\omega _{{\rm{ref}}}^{\mathrm{A}}}}&{{{\hat v}_{{\rm{ref}}}^{\mathrm{A}}}}&{{P_{{\rm{ref}}}^{\mathrm{B}}}}&{{Q_{{\rm{ref}}}^{\mathrm{B}}}}&{{\omega _{{\rm{ref}}}^{\mathrm{B}}}}&{{{\hat v}_{{\rm{ref}}}^{\mathrm{B}}}}&{{P_{{\rm{ref}}}^{\mathrm{C}}}}&{{Q_{{\rm{ref}}}^{\mathrm{C}}}}&{{\omega _{{\rm{ref}}}^{\mathrm{C}}}}&{{{\hat v}_{{\rm{ref}}}^{\mathrm{C}}}}
		\end{array}} \right]^ \top }
\end{equation}

\section{Numerical implementation and validation of the methodology}\label{sec:four}

In this section we describe how the singular perturbation theory is applied to the scenario modeled in Section~\ref{sec:three} for obtaining two sub-systems of reduced order which are completely decoupled. Our final purpose is to demonstrate the validity of the proposed methodology by means of time-domain simulations under different perturbations.

\subsection{Application of SP theory to the modeled scenario}
According to the discussion in Appendix \ref{sec:app}, a preliminary step for this model reduction is an accurate analysis of the time-scale evolution of the state variables. In particular, a first guess of which will be the fast and slow states can be obtained by looking at the eigenvalues of the linearization of our system around some given operation point: the fast states are those associated with large eigenvalues and the slow states correspond to small eigenvalues. At this regard, we have to stress that in our setting, with large and small we refer to the modulus of the (complex) eigenvalues.

Hence, starting from our original non-linear system $\dot{x} = f(x,u)$ we look for the equilibrium points $({\bar x},{\bar u})$ that are solutions of the algebraic equation $f({\bar x},{\bar u}) = 0$. With this equilibrium point we derive the following small signal model for the small-signal variation $(x_s,u_s)=(x-{\bar x},u-{\bar u})$:
\begin{align}\label{eq:linearized_system}
\dot{x}_s = Ax_s + Bu_s+g(x_s,u_s)
\end{align}
where $g$ denotes the high-order terms and the subscript $s$ indicates that we are now working with small-signal variables.

\begin{figure}[h!]
	\centering
	\includegraphics[width=0.6\textwidth]{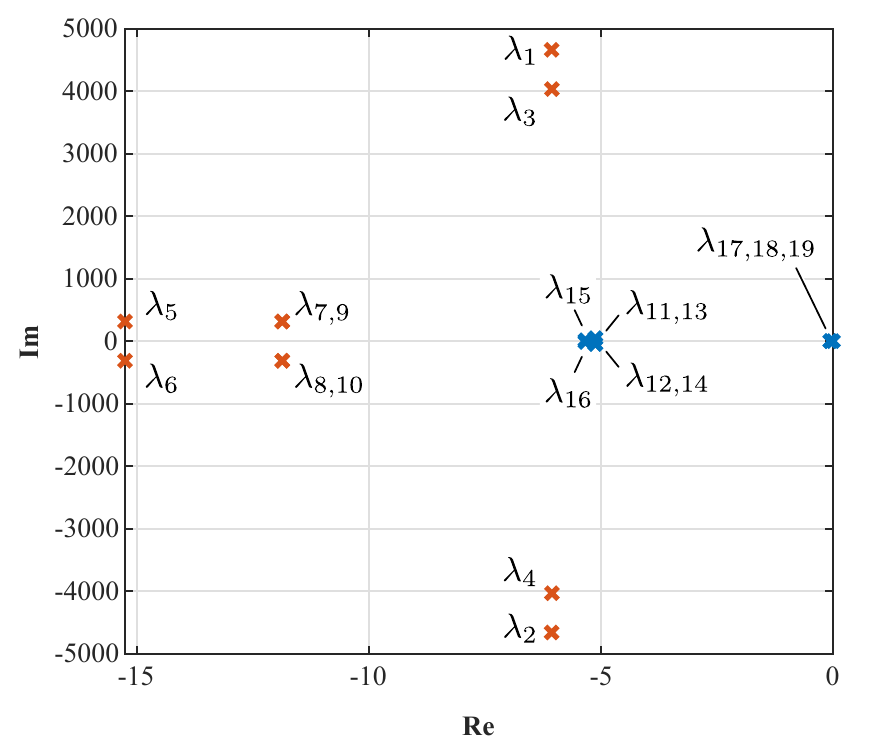}
	\caption{Eigenvalues of the modeled power system scenario}
	\label{fig:eigs}
\end{figure}

From this small-signal model we can compute the eigenvalues of the matrix $A$ in \eqref{eq:linearized_system} and select the largest ones. Figure~\ref{fig:eigs} illustrates the eigenvalues of the modeled power system for the parameters listed in Table~\ref{tab:params}.

% The parameters employed in the modeled scenario are summarized in Table~\ref{tab:params}.

\begin{table}[h!]
	\centering
	\caption{Parameters and inputs employed in the analysis}
	\label{tab:params}
	% \resizebox{0.5\textwidth}{!}{%
	\begin{tabular}{lclc}
		\toprule
		\textbf{Parameter} & \textbf{Value} & \multicolumn{1}{c}{\textbf{Parameter}} & \textbf{Value} \\ \midrule
		$S_b$ & \SI{2.75}{MVA} & $K_D$ & \num{1} \\
		$V_b$ & \SI{563}{\volt} & $T_a$ & \SI{2}{\second} \\
		$\omega_b$ & \SI{2\pi50}{\radian\per\second} & $V_\mathit{dc}$ & \num{1} p.u. \\
		$P_\mathrm{ref}$ & \num{0.5} p.u.   & $R_f$ & \num{0.003} p.u. \\
		$Q_\mathrm{ref}$ & \num{0} p.u. & $L_f$ & \num{0.08} p.u. \\
		${\hat{\bf{v}}}_{\mathrm{ref}}$ & \num{1} p.u. & $C_f$ & \num{0.074} p.u. \\
		$\omega_\mathrm{ref}$ & \num{1} p.u. & $V_g$ & \num{1} p.u. \\
		$K_\omega$ & \num{20} & $\omega_g$ & \num{1} p.u. \\
		$K$ & \num{2200} & $R_g$ & \num{0.01} p.u. \\
		$K_q$ & \num{0} & $L_g$ & \num{0.2} p.u. \\ \bottomrule
	\end{tabular}%
	% }
\end{table}

Figure~\ref{fig:eigs} depicts that in the linearized system the imaginary scale is much larger than the real one. The numerical values of the eigenvalues collected in Table~\ref{tab:eigs1} show that the fast eigenvalues (which are highlighted in red in Figure~\ref{fig:eigs}) exhibit a high imaginary part compared to the slow ones. In addition, these imaginary parts differ significantly from each other, meaning that the system contains states with different oscillation frequencies, which is expected in power systems with different line characteristics. 

\begin{table}[h!]
	\centering
	\caption{Numerical values and modulus of eigenvalues}
	\label{tab:eigs1}
	% \resizebox{0.5\textwidth}{!}{%
	\begin{tabular}{lccc}
		\toprule
		\textbf{Eig.} & \textbf{Value} & \textbf{Modulus} & \textbf{Type}\\ \midrule
		$\lambda _{1,2}$ & $-6.07\pm j4661.02$ & $4661.03$ & \multirow{5}{*}{\textit{Fast}}\\
		$\lambda _{3,4}$ & $-6.07\pm j4032.74$ & $4032.74$ & \\
		$\lambda _{5,6}$ & $-15.26\pm j314.16$ & $314.53$ & \\
		$\lambda _{7,8}$ & $-11.87\pm j314.17$ & $314.39$ & \\
		$\lambda _{9,10}$ & $-11.87\pm j314.17$ & $314.39$ & \\ \midrule
		$\lambda _{11,12}$ & $-5.13\pm j44.69$ & $44.96$ & \multirow{5}{*}{\textit{Slow}}\\
		$\lambda _{13,14}$ & $-5.13\pm j44.69$ & $44.96$ & \\
		$\lambda _{15,16}$ & $-5.34\pm j14.04$ & $15.03$ & \\
		$\lambda _{17,18}$ & $-0.0636$ & $0.064$ & \\
		$\lambda _{19}$ & $-0.0066$ & $0.0066$ & \\ \bottomrule
	\end{tabular}%
	% }
\end{table}

For completeness, we have to mention that the classical SP approach described, for instance, in \cite[Chapter 2]{kokotovic1999singular} would identify fast and slow states basing only on the real part of the eigenvalues. Nevertheless, as it has been pointed out in \cite{Kodra2016}, when applying SP to power systems involving highly oscillatory components this selection may yield to an inaccurate approximation of the overall dynamics. This is very relevant in the analyzed case, since there are several eigenvalues with very similar real parts but with a significant difference in their imaginary parts (e.g. $\lambda_1$--$\lambda_4$ and $\lambda_{11}$--$\lambda_{16}$). For this reason, in our case we select what we consider as large eigenvalues based on their modulus, thus also considering the imaginary part of the eigenvalue.

\begin{figure}[h!]
	\centering
	\includegraphics[width=0.6\textwidth]{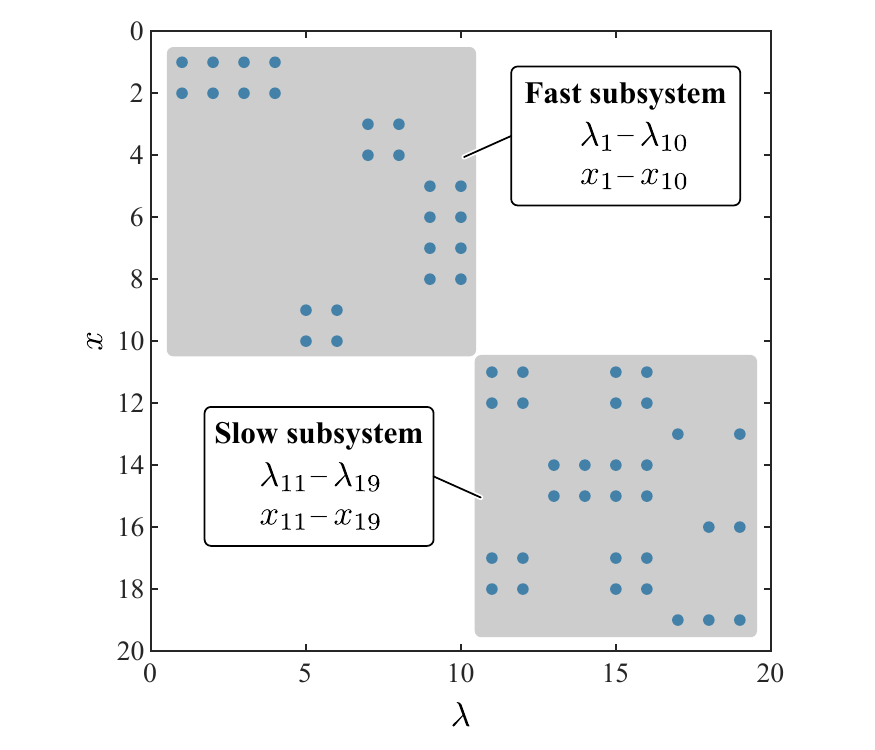}
	\caption{Most significant participation factors ($> 0.15$) of the modeled power system scenario}
	\label{fig:partfact1}
\end{figure}

Once we have identified the fast and slow eigenvalues it is fundamental to determine how they are related to the states of the system so that we can rearrange the equations \eqref{eq:linearized_system} in the form \eqref{eq:SP_general1}--\eqref{eq:SP_general2}. We have estimated this relation by calculating the weighted participation factors as in \cite{Shahnia2017}. Figure~\ref{fig:partfact1} illustrates with blue dots the weighted participation factors that are greater than $0.15$, meaning that the coupling between the state and the eigenvalue is significant.

These participation factors clearly show that the first ten eigenvalues ($\lambda _1$--$\lambda _{10}$), which are the fast eigenvalues, are directly coupled to the first ten states in the modeled scenario ($x _1$--$x _{10}$). Similarly, the eigenvalues $\lambda _{11}$--$\lambda _{19}$, which are the slow eigenvalues, are coupled to the states $x _{11}$--$x _{19}$ in the complete representation. This means that, for this particular case, the fast subsystem is representing the dynamics of the electrical part while the slow subsystem represents the dynamics of the inverter controllers.

Based on this separation of states we have carried out the computation of the matrix $L$ by finding a solution to equation \eqref{eq:L} through the following iterative scheme:
\begin{align}\label{eq:L_iter}
L_0=A_{22}^{-1}A_{21}, \quad L_{k+1}=A_{22}^{-1}A_{21}+A_{22}^{-1}L_k(A_{11}-A_{12}L_k), \quad k\geq 0,
\end{align}
which is exactly the same as the iterative method for the standard singular perturbation theory (see \cite[Chapter 2]{kokotovic1976singular}). 

In \eqref{eq:L_iter}, the matrices $A_{11}$, $A_{12}$, $A_{21}$, and $A_{22}$ are obtained from a block decomposition of the Jacobian of the original non-linear equations (the matrix $A$ in \eqref{eq:linearized_system}), according to the employed state separation. Then, if the eigenvalue condition \eqref{eq:eigenvalues} holds true, we can follow the procedure described previously to completely decouple the fast and slow states of the system.

\subsection{Time-domain simulations}
In order to demonstrate the validity of the proposed methodology, we compare the time-domain response of the original system with the results obtained with the decoupled subsystems.

The scenario has been configured with the parameters collected in Table~\ref{tab:params}, and the simulation begins from a steady-state point of operation to avoid any initial transient. At the instant $t=\SI{1}{\second}$ we modify the power reference of inverter $\mathrm{A}$ from $0.5$ p.u. to $0.6$ p.u. to compare the transient response of both systems.

\begin{figure}[h!]
	\centering
	\includegraphics[width=0.6\textwidth]{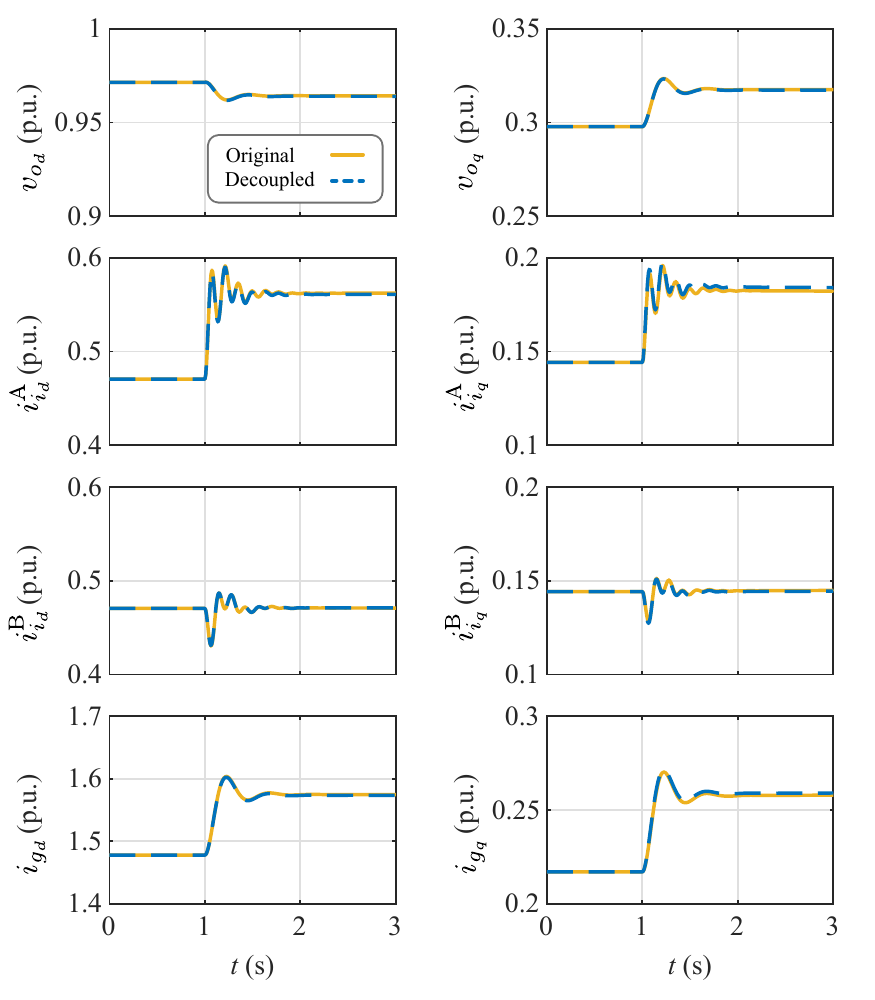}
	\caption{Transient response of the states associated to the fast eigenvalues of the modeled scenario}
	\label{fig:resultselec1}
\end{figure}

Figure~\ref{fig:resultselec1} illustrates some of the state variables associated to the electrical part of the scenario. In this case we show the dynamic behavior of the output voltage ($\mathbf{v}_o$), the output currents of inverter $\mathrm{A}$ and $\mathrm{B}$ ($\mathbf{i}_{i}^A$ and $\mathbf{i}_{i}^B$, respectively) and the grid-side current ($\mathbf{i}_{g}$). As we modify only the reference of inverter $\mathrm{A}$, the curves of inverter $\mathrm{B}$ and $\mathrm{C}$ are equal and therefore they have not been included in the figure.

\begin{figure}[h!]
	\centering
	\includegraphics[width=0.6\textwidth]{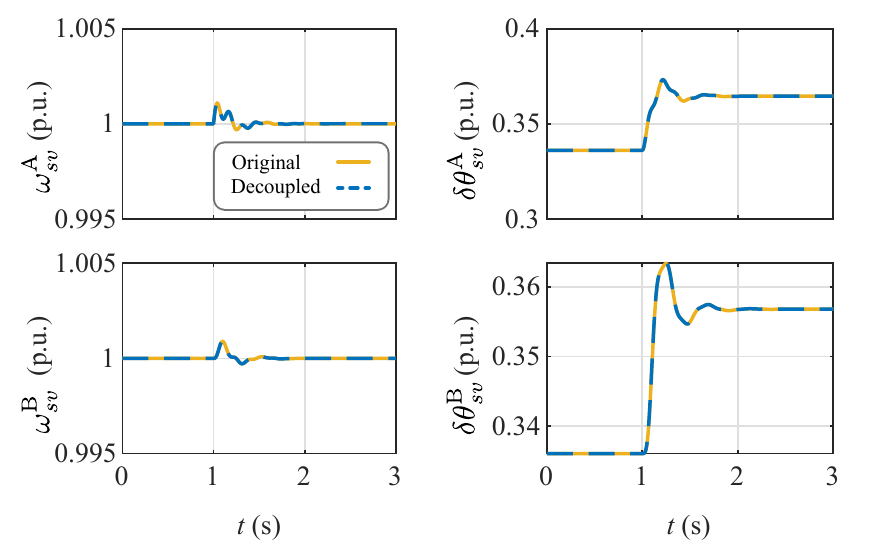}
	\caption{Transient response of the states associated to the slow eigenvalues of the modeled scenario}
	\label{fig:resultscont1}
\end{figure}

Similarly, Figure~\ref{fig:resultscont1} shows the time-domain results of the states of the inverter controllers. The variables shown are the controller frequencies and angle deviations of inverters $\mathrm{A}$ and $\mathrm{B}$ ($\omega_{sv}^A$ and $\omega_{sv}^B$, and $\delta\theta_{sv}^A$ and $\delta\theta_{sv}^B$, respectively).

The time-domain simulations illustrate how, when the reference of inverter $\mathrm{A}$ is increased at the instant $t=\SI{1}{\second}$, the \textit{d}-axis current at the grid-side increases smoothly from approximately $1.5$ p.u. to $1.6$ p.u., which corresponds to the sum of the currents provided by the three converters in parallel. Similarly, the \textit{d}-axis voltage decreases and the \textit{q}-axis voltage increases smoothly after this transient. The currents of the inverter $\mathrm{A}$ and $\mathrm{B}$ exhibit a fast oscillation superimposed to the smooth transient observed in the grid-side current. The main reason for this oscillation are the interactions between the three inverters, because inverters $\mathrm{B}$ and $\mathrm{C}$ control the output current to remain at their reference value. Regarding the controller states, we can observe that the oscillations in the controller frequencies are relatively small but the $\theta$ angles vary to adjust the power exchanged between the inverter and the grid.

In any case, the results in both figures demonstrate that the decoupled equations have a very high fidelity in comparison to the original equations even for a 20\% variation from the equilibrium point. The fast as well as slow dynamics of the original system are accurately represented by the decoupled system of equations obtained from the proposed methodology. This means that the original system of 19 ODEs can be simulated as two completely independent systems of 10 and 9 equations, and then the results can be converted back to the original variable notation by a simple matrix multiplication. This is interesting not only as a means to reduce the computational burden in the simulation of high-order dynamical systems, but it is also an efficient approach to simplify the equations to find candidate functions in which to apply the theorem of Lyapunov to determine the domain of attraction---and hence the stability region---of the system.

\section{Conclusions} \label{sec:conc}
There is an increasing demand to develop new tools and methods to model, analyze and control modern power systems in order to cope with the various time-scale oscillations caused by the massive integration of power electronic devices.

In this paper, we discuss the application of singular perturbation techniques for the model order reduction of this kind of power systems. In particular, we propose a general methodology which is applicable not only to electric power systems, but also to any dynamic system modeled with ordinary differential equations.

Our approach can be summarized in four points:
\begin{enumerate}
	\item To obtain the non-linear equations describing the time-evolution of the system.
	\item To linearize the model around some operation point and compute the eigenvalues and participation factors of the associated matrix.
	\item To select the largest (in terms of their modulus) eigenvalues and separate accordingly the states of the linearized system to compute two transformation matrices $L$ and $H$.
	\item To employ these matrices in suitable changes of variables in order to completely decouple the fast and slow states of the original non-linear system.
\end{enumerate}

In order to validate the proposed methodology, we have modeled a simplified power system scenario comprised by three inverters in parallel connected to a common grid. We have configured these converters to emulate the behavior of classical synchronous machines---i.e. as synchronverters---to contribute in the primary regulation and inertial response of the power system. Then, we have simulated this scenario for a variation in the power reference and the results have demonstrated that the decoupled system of equations obtained with the proposed methodology accurately represents the time-domain evolution of the dynamic states.

Besides, we remark that our MOR is carried out so that it maintains the non-linear nature of the original power system also in its final decoupled form. 

This means that the proposed approach is suitable not only to decrease the computational burden of highly complex dynamic systems, but also to obtain simplified and decoupled---yet non-linear---representations that can facilitate further analyses such as the application of Lyapunov techniques for the assessment of the stability or for non-linear control design.

\appendix

\section{Appendix} \label{sec:app}
\setcounter{equation}{0}
\renewcommand{\theequation}{A.\arabic{equation}}

\subsection{Mathematical background of Singular Perturbation theory}\label{app:sp}
This appendix is devoted to an abridged presentation of the mathematical background of the singular perturbation approach we employ in Section 4. This presentation is an adaptation of \cite[Chapter 2]{kokotovic1999singular} and \cite{Chow1978}.

The starting point is to rearrange the equations in our model, so that the ODE system takes the form:
\begin{subequations}
	\begin{align}
	\dot{z} = A_{11}z+A_{12}y+f(z,y), \quad z(t_0)=z_0,\label{eq:SP_general1}
	\\
	\dot{y} = A_{21}z+A_{22}y+g(z,y), \quad y(t_0)=y_0,\label{eq:SP_general2}
	\end{align}
\end{subequations}
with $A_{11}\in\mathbb{R}^{n\times n}$, $A_{12}\in\mathbb{R}^{n\times m}$, $A_{21}\in\mathbb{R}^{m\times n}$, $A_{22}\in\mathbb{R}^{m\times m}$, $m,n\in\mathbb{N}$ and where $f(z,y)$ and $g(z,y)$ indicate  high-order non-linear parts.

In the standard singular perturbation theory, equation \eqref{eq:SP_general2} would appear multiplied by a small parameter, which is often denoted by $\varepsilon$, indicating that these are the slow states of the system. In our case, this parameter is implicitly included in the system and our standing assumptions are to be able to extract the time-scale difference due to this hidden small parameter. This can be done if two conditions are satisfied:
\begin{enumerate}
	\item the matrix $A_{22}$ is non-singular, i.e. $\det(A_{22})\neq 0$, and that it is possible to find a matrix $L\in\mathbb{R}^{m\times n}$ such that the following equation holds:
	\begin{align}\label{eq:L}
	A_{21}-A_{22}L+LA_{11}-LA_{12}L = 0.
	\end{align}
	In practical applications, this matrix $L$ is usually obtained through the iterative scheme \eqref{eq:L_iter};
	\item Both $A_{11}-A_{12}L$ and $A_{22}+LA_{12}$ are stable matrices namely, eigenvalues of them lie in the open left half plane in the complex plane and all eigenvalues of $A_{22}+LA_{12}$ are simple complex numbers with nonzero imaginary parts. Moreover, it holds that 
	\begin{align}\label{eq:eigenvalues}
	\min\Big\{|\mathrm{Im}\lambda_j(A_{22}+LA_{12})|:j=1,\ldots,m\Big\}\gg\max\Big\{|\mathrm{Im}\lambda_j(A_{11}-A_{12}L)|:j=1,\ldots,n\Big\},
	\end{align}
	where $\mathrm{Im}$ represents the imaginary part of complex number. 
\end{enumerate}
We note that the situation described in the second assumption on the imaginary parts of subsystem eigenvalues is often observed in power systems with converters and this appendix describes how to separate the highly oscillatory modes from such a non-linear system based on \cite{Chow1978}. 

We first separate two time-scales in (\ref{eq:SP_general1})-(\ref{eq:SP_general2}) by the change of variables $\eta=y+Lz$ to get the so-called \textit{block-triangular form}, that is,
\begin{align}\label{eq:block_triangular}
\left[\begin{matrix} \dot{z} \\ \dot{\eta}\end{matrix}\right] = \left[\begin{matrix} A_{11}-A_{12}L & A_{12} \\ 0 & A_{22}+LA_{12}\end{matrix}\right]\left[\begin{matrix} z \\ \eta\end{matrix}\right] + \left[\begin{matrix} f(z,\eta-Lz) \\ \tilde{g}(x,\eta-Lz)\end{matrix}\right], \quad \left[\begin{matrix} z(t_0) \\ \eta(t_0)\end{matrix}\right] = \left[\begin{matrix} z_0 \\ y_0+Lz_0\end{matrix}\right], 
\end{align}
where $\tilde{g}(z,y):=Lf(z,y)+g(z,y)$.

From the second condition, we can rewrite $A_{22}+LA_{12}$ as 
\[
T(A_{22}+LA_{12})T^{-1}= \bmat{ D_1 & D_2/\varepsilon \\ D_3/\varepsilon & D_4},
\]
where small parameter $\varepsilon>0$ is introduced from (\ref{eq:eigenvalues}) and $T\in\mathbb{R}^{m\times m}$ is a suitable nonsingular matrix. Typically, it can be seen that $D_2$, $D_3$ take the forms $D_2=\mathrm{diag}(-\omega_1,\ldots,-\omega_m)$,  $D_3=\mathrm{diag}(\omega_1,\ldots,\omega_m)$ where $\omega_j\in\mathbb{R}$, $j=1,\ldots,m$. Introducing a notation 
\[
D(\varepsilon):= \bmat{\varepsilon D_1 &D_2\\ D_3&\varepsilon D_4},
\]
the equation for $\eta$ in (\ref{eq:block_triangular}) is equivalent to 
\[
\varepsilon \dot{\eta} =T^{-1}D(\varepsilon)T\eta +\varepsilon\bar{g}(x,\eta-Lz).
\]

We, next, apply another linear transformation, $\xi = x- H\eta$ with $H\in\mathbb{R}^{n\times m}$ solution of the Sylvester equation 
\begin{align}\label{eq:sylvester}
(A_{11}-A_{12}L)H-H(A_{22}+LA_{12})+A_{12} = 0,
\end{align}
and we obtain from \eqref{eq:block_triangular} the diagonal system
\begin{align}\label{eq:block_diagonal}
&\left[\begin{matrix} \dot{\xi} \\ \dot{\eta}\end{matrix}\right] = \left[\begin{matrix} A_{11}-A_{12}L & 0 \\ 0 & A_{22}+LA_{12}\end{matrix}\right]\left[\begin{matrix} \xi \\ \eta\end{matrix}\right] + \left[\begin{matrix} \bar{f}(\xi+H\eta,(I-LH)\eta-L\xi) \\ \bar{g}(\xi+H\eta,(I-LH)\eta-L\xi)\end{matrix}\right], \notag 
\\[8pt]
&\left[\begin{matrix} \xi(t_0) \\ \eta(t_0)\end{matrix}\right] = \left[\begin{matrix} (I-HL)z_0-Hy_0 \\ y_0+Lz_0\end{matrix}\right],
\end{align}
where $\bar{f}(z,y)=(I-HL)f(z,y)-Hg(z,y)$. 
We remark that the existence of a matrix $H$ satisfying the Sylvester equation \eqref{eq:sylvester} is a consequence of the eigenvalue condition \eqref{eq:eigenvalues}. 

System (\ref{eq:block_diagonal}) is equivalently rewritten in a non-linear singular perturbation form:
\begin{subequations}\label{eq:xi-eta-epsilon}
	\begin{gather}
	\dot\xi =(A_{11}-A_{12}L)\xi +\tilde{f}(\xi,\eta)\\
	\varepsilon\dot\eta= T^{-1}D(\varepsilon)T\eta +\varepsilon\tilde{g}(\xi,\eta),
	\end{gather}
\end{subequations}
where $\tilde{f}(\xi,\eta)=\bar{f}(\xi+H\eta,(I-LH)\eta-L\xi)$, 
$\tilde{g}(\xi,\eta)=\bar{g}(\xi+H\eta,(I-LH)\eta-L\xi)$. However, the meaning of the singular perturbation is different from the standard theory because eigenvalues of $D(\varepsilon)/\varepsilon$ approach infinity along asymptotes that are parallel to the imaginary axis. The decomposition, therefore, should be carried out based on variation of signals rather than slow and fast attenuation. We say that a solution $q(t,\varepsilon)$ of (\ref{eq:xi-eta-epsilon}) is slowly varying if there exist constants $\varepsilon^\ast>0$ and $C>0$ independent of $\varepsilon$ such that $|\dot q(t,\varepsilon)|<C$ for all $t\geqslant t_0$ and $\varepsilon\in(0,\varepsilon^\ast)$.
It can be shown that $\eta(t)$ does not contain slowly varying parts as follows. By the variation of constants formula, we have 
\[
\eta(t)=\Phi(t,t_0)\eta(0)+\int_{t_0}^t\Phi(t,s)\tilde{g}(\xi(s),\eta(s))\,ds, 
\]
where $\Phi(t,s)=\exp\left[ \frac{T^{-1}D(\varepsilon)T}{\varepsilon}(t-s) \right]$ and therefore, we see that $\dot\eta$ contains the term $\frac{T^{-1}D(\varepsilon)T}{\varepsilon}\Phi(t,0)\eta(0)$. Noting that entries in $\Phi(t,t_0)$ are of the form 
\[
\exp(-a_j(t-t_0))\sin(\omega_j(t-t_0)/\varepsilon), 
\quad \exp(-a_j(t-t_0))\cos(\omega_j(t-t_0)/\varepsilon)
\]
and therefore $\Phi(t,t_0)$ is bounded as $\varepsilon\to+0$, $\dot\eta$ is unbounded as $\varepsilon\to+0$. The slowly varying solution of (\ref{eq:xi-eta-epsilon}) is obtained from 
\[
\dot{\bar{\xi}}=(A_{11}-A_{12}L)\bar{\xi}+\tilde{f}(\bar{\xi},0)
\]
and $\xi(t)$ is approximated to $O(\varepsilon)$ by $\bar{\xi}(t)$. The oscillatory part $\eta(t)$ is described by the time-scale of $\tau=t/\varepsilon$, in which $\xi(t)$ can be considered constant. 
In view of that, the equations describing slowly varying part and highly oscillatory part separately can be obtained approximately as 
\begin{align*}
&\dot{\xi}=(A_{11}-A_{12}L)\xi +\tilde{f}(\xi,0)
\\
&\dot{\eta}=(A_{22}+LA_{12})\eta  +\tilde{g}(\xi(t_0),\eta).
\end{align*}

% References
\bibliographystyle{myIEEEtran_noURL}  % myIEEEtran_noURL

\bibliography{biblio}\ %IEEEabrv instead of IEEEfull

\end{document}